# JConstHide: A Framework for Java Source Code Constant Hiding

Praveen Sivadasan[1], P Sojan Lal[2]

[1]School of Computer Sciences,
Mahatma Gandhi University, Kerala, India,
*praveen_sivadas@yahoo.com*

[2]School of Computer Sciences,
Mahatma Gandhi University, Kerala, India,
*sojanlal@gmail.com*

*Abstract*— Software obfuscation or obscuring a software is an approach to defeat the practice of reverse engineering a software for using its functionality illegally in the development of another software. Java applications are more amenable to reverse engineering and re-engineering attacks through methods such as decompilation because Java class files store the program in a semi complied form called 'byte' codes. The existing obfuscation systems obfuscate the Java class files. Obfuscated source code produce obfuscated byte codes and hence two level obfuscation (source code and byte code level) of the program makes it more resilient to reverse engineering attacks. . But source code obfuscation is much more difficult due to richer set of programming constructs and the scope of the different variables used in the program and only very little progress has been made on this front. We in this paper are proposing a framework named 'JConstHide' for hiding constants, especially integers in the java source codes, to defeat reverse engineering through decompilation. To the best of our knowledge, no data hiding software are available for java source code constant hiding.

## I. INTRODUCTION

The java based web applications gained popularity because of its Architecture Neutral Distribution Format (ANDF). During compilation, the Java source code is translated to java class files that contain Java Virtual Machine (JVM) code called the 'byte code', retaining most or all information present in the original source code. This is because the translation to real machine instruction happens in the browser of the user's machine by JIT (Just-In-Time Compiler). Also, Java programs are small in size because of the vast functionalities provided by the Java standard libraries. Decompilation is the process of generating source codes from machine codes or intermediate byte codes. Though decompilation is in general hard for most programming languages, the semi compiled nature of Java class files makes it more amenable to reverse engineering [2] and re-engineering attacks through decompilation. Reverse engineering can be defined as the process of analyzing a subject system to 1) identify the system's components and their interrelationships, and, 2) create representations of the system at higher levels of abstraction. Reverse engineering involves the extraction of design elements from an existing system, but it does not involve modifying the target system or generating new systems. Reengineering is the modification of a software system that takes place after it has been reverse engineered, generally to add new functionality, or to correct errors.

This makes it easier for the competitors to extract the proprietary algorithms and data structures from Java applications in order to incorporate them into their own programs in order to cut down their development time and cost. Such cases of intellectual property thefts [6, 7, 8] are difficult to detect and pursue legally. Statistics [5] show that four out of every ten software programs is pirated worldwide and over the years, global software piracy has increased by over 40% and has caused a loss of more than 11 billion USD. Software obfuscation [1,6] is a popular approach where the program is transformed into an obfuscated program using an 'obfuscator'[3] in such a way that the functionality and the input/output behavior is preserved in the obfuscated program whereas it is much more difficult to reverse engineer the obfuscated program. The obfuscation can be preformed on the source code, the intermediate code or the machine executable code. Data transformation [1,4] and constant hiding are the two well studied obfuscation techniques and the tool is based on constant hiding technique which is discussed in the following section.

In [5], Ertaul et. al proposed a novel constant hiding techniques using *y-factors*. The y-factors are essentially a predefined increasing sequence of 'm' prime numbers y[0], y[1],y[2]...,y[m]. The y_factors can be used to transform a non negative number 'x' which is less than y[0] as follows. Let the function 'F(A, k)' be defined as F(A, k) = ((....((A mod y[k]) mod y[k-1]) mod y[k-2]) .... mod y[0]). Now replace 'x' by the expression F(A, k) such that F(A, k) evaluates to 'x'. Now to hide any large positive constant say 'c' in the program, first 'c' is replaced with a simple expression of the form 2*d + r where 'r' is 0 if 'x' is even and 'r' is 1 if 'x' is odd. Now, the constants 2 and 'r' in the resulting expression can be hidden by replacing it with the corresponding F( ) function.

### a) The ConstHide Module

To compute the function F( ) as defined in the introduction, we use an array Y[m] of 'm' pairs where Y[i] = (Pi, Qi) denote the pair at the i-th index of Y. These pairs have the following properties: a) for any pair Y[i] = (Pi, Qi), Pi + Qi is a prime number and b) if i < j then $P_i + Q_i < P_j + Q_j$. That is, sum of the numbers in any pair is a prime number and the pairs are stored in Y array in the increasing order of their sum value. The following sequence of pairs for example can be the contents of the Y-array -

```
(2,3),(5,6),(11,12),(23,24),(47,48),(95,96),(191
,192),….. (1287, 12288).
```

Function F( ) is computed using the following algorithm.

```
int F(A, k){
    //k is a number between 1 and m which
    //denotes the depth of the obfuscation.
    Y[m]={(P1,Q1),(P2,Q2)........(Pm,Qm)}
    r = A;
    for (i :k .....1) {r = r mod (Pi + Qi);}
    return r; }
```

The ConstHide would for example hide the constant '2' by replacing it with an expression chosen randomly from the list: `F(41%23,2)`, `F(374%191,5)`, `F(757%383,6)` and so on. Though most compilers simplify the expressions of the form 374%191, we still use these expressions to ensure that the source code itself is difficult to comprehend.

Now, let us discuss about the tool operation (Figure 1) where the tool parses the java source code and the output is a constant hidden source code. Again, the output file can be chosen to implement further levels of obscurity by data hiding.

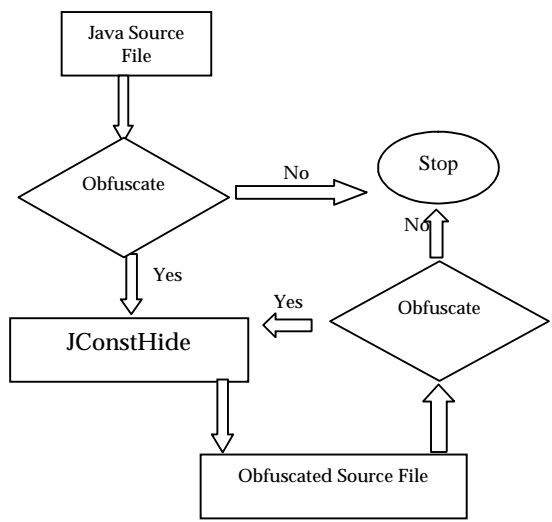

Figure 1. JConstHide Framework

The tool operation is detailed in the next section.

In Java programs, a statement terminates with a ; symbol. On tool invocation, initially the tool parses each source code statement for symbols ; and End of Line (EOL). A new file is rewritten for the source code where each statement terminates with a ';' or an EOL. Followed by that, the tool parses each java source code statement of the new file for tokens like [ , = ( + / * - space < > % and thereby replaces the first constant immediately followed by any of the tokens, by F (a, b) call where the function definition is included in the file 'obfuscate.java'. We call the tool output file as 'obfuscated file' and on selection of the obfuscated file, the tool further parses for F(a, b) call, and the first constant, say 'h' of argument 'a' of F() will be transformed into '2*d + r' with the constant 2 hidden by the F() call and for each successive iteration, this process is again repeated for the first argument of the recursive F() calls.

We shall explain the tool operation by considering a snippet of source code 'leapyears.java',

```
class leapyears
{
    public static void main(String[] args)
    {
int i=2006;int n;for (n=1990; n<=i ; n++){
int l=n%4;if (l==0){
    System.out.println("leap year: "+n);
                }}}}
```
Figure2. Java code to find Leap years between 1990 and 2006

The first iteration of obfuscation on the snippet, generates an output file where statements terminate with ';' or an End of Line(EOL), as shown below

```
public class leapyears_mod
{
public static void main(String[] args)
{
int i=2006;
int n;
for (n=1990;
 n<=i ;
 n++){
    int l=n+4;
    if (l==0){
    System.out.println("leap year: "+n);
                }}}}
```

Figure3. Tool output after the first iteration of obfuscation

The second iteration of obfuscation, replaces the first constant of the statement immediately followed by any of the tokens [ , = ( + / * - space < > % by F(a,b) call.

```
public class leapyears_mod123 extends obfuscate
  {
  public static void main(String[] args)
  {
int i=(1003*F(12273%6143,10));
int n;
for (n=(995*F(757%383,6));
n<=i ;
 n++){
int l=n%(2*F(49135%24575,12));
if (l==(0*F(374%191,5))){
System.out.println("leap year: "+n);
    }}}}
```
Figure4. Tool output after the second iteration

The next iteration of obfuscation, parses for the first constant of F(a,b) call and replaces it by F( ) call, resulting in recursive F() calls.

```
public   class   leapyears_mod123123      extends
obfuscate
{
public static void main(String[] args)
{
inti=(1003*F((F(3059%1535,8)*6136+F(33%21,2))%61
43,10));
int n;
for(n=(995*F((F(49135%24575,12)*378+F(33%21,2))%
383,6));
n<=i ;
n++){
intl=n%(2*F((F(757%383,6)*24567+F(33%21,2))%2457
5,12));
if (l==(0*F(F(1524%767,7)*187%191,5))){
System.out.println("leap year: "+n);
}}}}
```
Figure5. Tool output after the third iteration

Hence, we can see that the various obfuscated codes results in recursive function calls leading to a possibility of more execution time. To analyse the execution time of the obfuscated codes, a program 'search_random.java' and its obfuscated versions are analysed. The algorithm of the code is as follows,

*Initialize array 'A' of size 100000, to 0*
*Read n*
```
for (i : 0 .... n-1){
```
*Generate a random number say 'num'<n*
*if((num%2)==0) Access A[num] else Set A[num] }*

For 100000 elements, the execution time analysis of the above code is performed on a system with Intel Core Duo processor, 1.66GHz, with 1GB of RAM and with Windows XP Professional operating system. Let C represent the original code and C1 represent the newly formatted non obfuscated version of the source code. Let C2, C3, C4 represent the different successive obfuscated versions of C, obfuscated by data hiding.

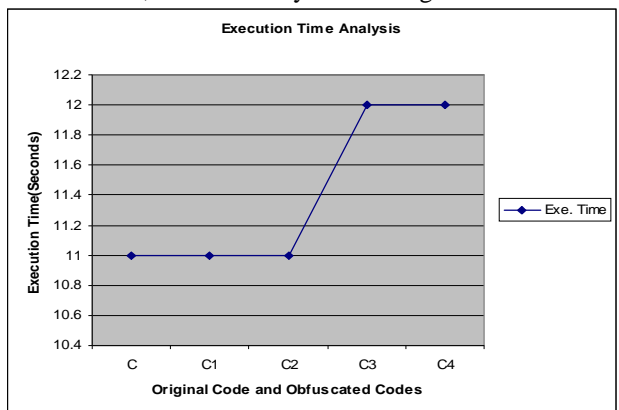

Figure6. Execution time analysis of a code and its obfuscated versions

The plot clearly reveals that the obfuscated codes are not having much deviation in execution times as with that of the original code.

Reverse engineering effort depends on the effort for comprehending entire code statements and is proportional to the number of statements. Hence, the repeated operation on the obfuscated code by the tool, adds more reverse engineering effort to the statements by generating recursive F() calls.

Due to the lack of commercial de-obfuscators in the market, this analysis of the obfuscated codes is solely based on its decompiled codes. The decompiled codes for the above obfuscated codes have been analyzed using FrontEnd Plus v1.04 decompiler and the decompiled code for the various iterated versions also contains the same number of recursive F() calls making the decompiled code hard to reverse engineer.

| Decompiled Codes | Number of Statements | No. of Obfuscated Statements |
|---|---|---|
| Leapyears.java | 6 | 4 |
| search_random.java | 26 | 13 |

Table 1. Analysis of decompiled codes

Let 'S' be the minimum number of statements of F ( ) call and 'N' be the number of successive iterations for obfuscation, then the reverse engineering effort 'RE', contributed by the final obfuscated version is given by RE=S*N obfuscated statements.

The Table1 shows that only statements involving constants are obfuscated. Therefore, we infer that the reverse engineering effort grows linearly for different iterations, not causing too much cost on execution time. Hence, the tool cuts down reverse engineering on the codes involving constants to an extent by adding more reverse engineering effort and the tool is found to be ineffective on codes without considerable constants.